# Discretized conical waves in multimode optical fibers

Bertrand Kibler and Pierre Béjot [*]

*Laboratoire Interdisciplinaire Carnot de Bourgogne, UMR6303 CNRS-UBFC, 21000 Dijon, France*

Multimode optical fibers has emerged as the platform that will bridge the gap between nonlinear optics in bulk media and in single-mode fibers. However, the understanding of the transition between these two research fields still remains incomplete despite numerous investigations of intermodal nonlinear phenomena and spatiotemporal coupling. Some of the striking phenomena observed in bulk media with ultrashort and ultra-intense pulses (i.e., conical emission, harmonic generation and light bullets) require a deeper insight to be possibly unveiled in multimode fibers. Here we generalize the concept of conical waves described in bulk media towards structured media, such as multimode optical fibers, in which only a discrete and finite number of modes can propagate. Such propagation-invariant optical wave packets can be linearly generated, in the limit of superposed monochromatic fields, by shaping their spatiotemporal spectrum, while they can spontaneously emerge when a rather intense short pulse propagates nonlinearly in a multimode waveguide, whatever the dispersion regime and waveguide geometry. The modal distribution of optical fibers then provides a discretization of conical emission (e.g., discretized X-waves). Future experiments in commercially-available multimode fibers could reveal different forms of conical emission and supercontinuum light bullets.
DOI:

*Introduction.* The field of nonlinear fiber optics has remained a growing area of research since the end of 1970s, boosted each decade by new technological advances [1]. In parallel, nonlinear optics in bulk media has attracted a great deal of attention, through the emergence of intense femtosecond lasers. High powers delivered by such laser sources have allowed to observe a large range of nonlinear optical phenomena in bulk materials [2-5], while single-mode fibers have enabled nonlinear effects through better light confinement and longer propagation distance [1,6]. In both fields, a strong interest has been manifested in the study of supercontinuum generation taking place during nonlinear pulse propagation [7,8], thus illustrating the connection between the developments of nonlinear optics in fibers and bulks.

Despite some similarities, single-mode fibers and bulk media have deep intrinsic differences. While an infinite and dense number of transversal modes can propagate in a bulk medium, only one exists in single-mode fibers. Recently, the revival of multimode fibers (MMFs) has led to investigating intermodal nonlinear phenomena and spatiotemporal couplings, mainly associated with broadband frequency conversion processes [9-10]. As several (discretized) modes propagate in MMFs, such waveguides appear as a relevant intermediate platform.

In this context, one can wonder whether spatiotemporal dynamics observed in bulk media can be transposed to the case of MMFs. For instance, conical emission is one of the striking phenomena observed in bulk samples or condensed media [4]. It is worth to underline that conical emission turns out to be a particular example of conical waves [4-5]. Such fully localized waves have been then described as non-dispersive and non-diffractive with propagation due to their intrinsic coupled spatiotemporal properties, mainly driven by a phase-matching condition defined by the bulk material dispersion. More generally, space-time wave packets in free space with arbitrary group velocities has recently focused a lot of attention [11-12]. The unique features of such fields that can be linearly shaped open new ways of controlling light–matter interactions.

In this work, we theoretically and numerically explore the concept of discretized conical waves and their properties in multimode fibers. Such optical wave packets result from a linear superposition of fiber modes with an engineered spatiotemporal spectrum. They can also spontaneously emerge during nonlinear propagation of ultrashort pulses similarly to bulk conical emission.

*Theoretical description.* Conical waves are wave packets localized both in space and time which sustain their spatiotemporal shapes during propagation in a bulk medium [4,13]. This property results from their intrinsically coupled spatiotemporal structures by which the combined effects of spatial diffraction and temporal dispersion cancel each other. In particular, conical waves have been shown to spontaneously emerge during the nonlinear propagation of ultrashort and ultra-intense laser pulses in bulk media [4,14-17]. In general, conical emission observed during the filamentation process is nothing but the manifestation of X-wave generation. Conical waves can either take the form of a hyperbola (X-wave) or an ellipse (O-wave), as well as their combination (Fish-wave), in the $(k_\perp, \omega)$ plane [18-20], where $k_\perp$ is the transversal component of the wave vector and $\omega$ the frequency ($\omega_0$ the carrier frequency). This usually depends on pulse propagating in the normal or the anomalous dispersion regime. While a dense number of modes can propagate within a bulk medium, structured and/or finite media (i.e., waveguides) support only a discrete number of modes at a given frequency $\omega$ that can propagate within the waveguide core. Note however, that the combination of core-guided, clad-guided and evanescent modes forms an (infinite) orthogonal basis set in which any electric field $E$ can be decomposed as follows:

$$E(r,t,z) = \int \sum_{m=1}^{\infty} \bar{E}(m,\omega)\mathcal{F}(r,m,\omega)e^{-i\omega t}d\omega \quad (1)$$

where $\mathcal{F}(r,m,\omega)$ is the transversal shape of the mode $m$ for a given $\omega$.

The unidirectional pulse propagation equation generalized to structured media (i.e., media embedding a transversal distribution of the refractive index) expressed in the modal basis reads in the linear regime as:

$$\partial_z \bar{E}(m,\omega) = iK_z(m,\omega)\bar{E}(m,\omega) \quad (2)$$

where $\bar{E}(m,\omega)$ is the electric field coordinate on the mode $m$ at a frequency $\omega$ and $K_z$ is the propagation constant of the mode $m$ at a given $\omega$. Writing the electric field $E(r,t,z) = A(r,t,z)e^{i(K_{z(1,\omega_0)}^{(0)}z - \omega_0 t)}$ and considering a frame propagating at the group velocity $1/K_{z(1,\omega_0)}^{(1)}$, where $K_{z(1,\omega_0)}^{(0)} = K_z(1,\omega_0)$ and $K_{z(1,\omega_0)}^{(1)} = \{\partial_\omega K_z(1,\omega)\}_{\omega=\omega_0}$, one obtains the following unidirectional pulse propagation equation:

$$\partial_z \bar{A}(m,\omega-\omega_0) = i\big[K_z(m,\omega-\omega_0) - K_{z(1,\omega_0)}^{(0)} - K_{z(1,\omega_0)}^{(1)}(\omega-\omega_0)\big]\bar{A}(m,\omega-\omega_0) \quad (3)$$

One can define families of modes that all satisfy the same relation:

$$K_z(m,\omega-\omega_0) - K_{z(1,\omega_0)}^{(0)} - K_{z(1,\omega_0)}^{(1)}(\omega-\omega_0) = \delta K_z^{(0)} - \delta K_z^{(1)}(\omega-\omega_0) \quad (4)$$

where $\delta K_z^{(0)}$ and $\delta K_z^{(1)}$ can be arbitrarily chosen constants. Any superposition of modes belonging to a given family forms a wave packet localized in both space and time, propagating without any distortion at a chosen group velocity $1/\big(K_{z(1,\omega_0)}^{(1)} + \delta K_z^{(1)}\big)$. Note that Eq. (4) is also valid for bulks. However, in the latter case, at any $\omega$ corresponds a value of $k_\perp$ (since the latter can take any value) for a given family of conical waves [21]. This is not true anymore in the context of discretized conical waves generated in waveguides. Indeed, the family of transversal modes propagating in the waveguide core is discrete. As a result, only a discrete number of frequencies (denoted hereafter $\Omega_m = \omega_m - \omega_0$) satisfy the relation (4).

The electric field component of a discretized conical wave is then composed of delta functions:

$$\bar{E}(m,\omega) = \bar{\epsilon}(m)\delta(\omega-\omega_m) \quad (5)$$

In the temporal domain, the envelope of a discretized conical wave then reads:

$$A(r,t) = \sum_{m=1}^{\infty} \bar{\epsilon}(m)\mathcal{F}(r,m,\omega_m)e^{-i\Omega_m t} \quad (6)$$

Since the conical wave follows Eq. (4), its complex amplitude will evolve during the propagation as:

$$A(r,t,z) = \sum_{m=1}^{\infty} \bar{\epsilon}(m) \mathcal{F}(r,m,\omega_m)e^{-i[\Omega_m t - (\delta K_z^{(0)} + \delta K_z^{(1)}\Omega_m)z]} \quad (7)$$

The resulting space-time wave-packet formed by the linear superposition of monochromatic modes is not impacted by the propagation; its invariant intensity shape directly depends on the phase-matched frequencies from Eq. (4) and waveguide properties. In Fig. 1(a), an example of discretized conical wave is numerically constructed in the space-time domain, by using the MMF parameters studied in the next section (with $\delta K_z^{(1)} = 22$ ps m$^{-1}$ and $\delta K_z^{(0)} = 0$). In particular, we considered the X-shaped pattern of the field distribution in the $(m,\omega)$ domain (see red squares in Fig. 1(b)), when pumping in the normal dispersion regime. In practice, by considering a limited number of narrowband modes, such configurations of conical waves could be generated based on usual wave shaping techniques with a spatial mode multiplexer. Here the corresponding $(r,t)$ pattern is characterized by a quasi-periodicity and a central V-shaped structure that exhibits a peak in the radial center (see Fig. 1(a)). Two symmetric decaying tails define a spatiotemporal cone vanishing at the core-clad interface of the waveguide.

By contrast, all modes involved could rather have a non-zero spectral width in the case of spontaneous emergence of conical waves during nonlinear propagation of ultrashort pulses (see next section). Then, a realistic nonlinearly-induced conical wave will write as Eq. (1) with $\bar{E}(m,\omega)$ a relatively narrow function centered on $\omega_m$ representing the spectral complex amplitude of the conical wave component whose carrier frequency oscillates at $\omega_m$ in the $m^{\text{th}}$ mode. As a consequence, the temporal complex amplitude $A(r,t)$ of a conical wave can be well approximated by

$$A(r,t) = \sum_{m=1}^{\infty} a_m(t)\mathcal{F}(r,m,\omega_m)e^{-i\Omega_m t} \quad (8)$$

where $a_m(t)$ is the temporal envelope of the component in the $m^{\text{th}}$ mode at a carrier frequency $\omega_m$. Accordingly, its complex amplitude will evolve during the propagation as:

$$A(r,t,z) = \sum_{m=1}^{\infty} a_m(t - (K^{(1)}_{z(m,\omega_m)} - K^{(1)}_{z(1,\omega_0)})z)$$
$$\mathcal{F}(r,m,\omega_m)e^{-i[\Omega_m t - (\delta K_z^{(0)} + \delta K_z^{(1)}\Omega_m)z]} \quad (9)$$

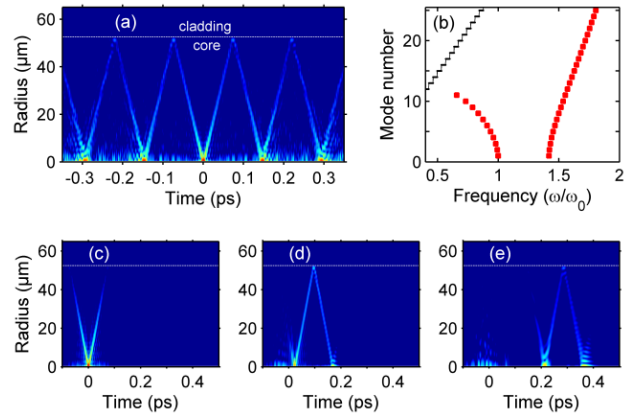

FIG. 1. Example of conical waves in a step-index MMF. (a) Full spatiotemporal power (over 3 decades). (b) Modal distribution of phase-matched frequencies (red squares). Black curve: number of guided modes. (c-e) Evolution of a realistic (i.e., dispersive) ultrashort (80-fs duration components) conical wave with propagation distance (for $z = 0$, 1, and 3 mm, respectively).

Then, the temporal envelope of the modes composing the conical wave (i.e., a broadband ultrashort pulse structure) does not propagate at the same group velocity. The propagation leads to a temporal dispersion of the conical wave, as depicted in Fig. 1(c-e), which depends on the spectral width of each component.

*Numerical simulations and results.* Our numerical approach of nonlinear pulse propagation in MMFs is based on the multimode unidirectional pulse propagation equation recently derived in Ref. [22], which describes the evolution of the complex electric field in the scalar approximation:

$$\partial_z \bar{A} = i\left[K_z - K^{(1)}_{z(1,\omega_0)}(\omega - \omega_0)\right]\bar{A} + \frac{in_0 n_2 \omega^2}{c^2 K_z}$$
$$\{(1-f_R)\overline{|A|^2 A} + f_R\overline{[\int h_R(\tau)|A(t-\tau)|^2 d\tau]A}\}, \quad (10)$$

where $n_0$ is the core refractive index calculated at $\omega_0$, the pump frequency. $n_2 = 3.2 \times 10^{-20}$ m$^2$ W$^{-1}$ is the nonlinear refractive index of silica glass [1]. The function $h_R$ is the Raman response with fraction $f_R = 0.18$ for fused silica glass. The propagation equation is expressed in a retarded frame moving at velocity $1/K^{(1)}_{z(1,\omega_0)}$. We solve the propagation by a split-step algorithm as described in Ref. [22].

In the following, we compare the numerical results with our theoretical approach to unveil the spontaneous formation of discretized conical waves in MMFs, more specifically in the case of femtosecond pulse pumping and input peak powers around the critical self-focusing threshold of silica glass. We first investigate the nonlinear propagation of 100-fs Gaussian pulse injected into the fundamental mode of a 4-cm-long segment of step-index MMF. The pump pulse is fixed at $\lambda_0 = 800$ nm, in the normal dispersion regime of the fiber. The fiber under study exhibits typical features of commercially-available MMFs (pure silica core with diameter $\phi = 105$ μm, and numerical aperture $NA = 0.22$). Our scalar approach limited our studies to the linearly polarized modes of $LP_{0,n}$ class. As the considered fiber is circularly symmetric, the coupling coefficients are nonzero only for $LP_{0,n}$ modes, when pumping configuration is a beam coupling into the fundamental mode of the fiber (namely $LP_{0,1}$ mode). We recall that a single mode excitation and its stable propagation over meter-long segments is possible in MMFs [23-24].

Figure 2(a) shows the distribution of the full optical spectrum (power in log. scale) over the different fiber modes after 4 cm of propagation. We observe here an evident X-shaped pattern of the field distribution that is very similar to one typical example of conical waves (X waves) well studied in the Fourier $(k_\perp, \omega)$ domain for bulk media [4]. The spontaneous formation of this energy spreading in higher-order modes from the fundamental mode can be retrieved by using the theoretical relation (4) of discretized conical waves (see white squares). We confirm the formation of a discretized conical wave for which the blue and highest frequencies are continuously contained in higher-order modes. Regarding the interpretation of the pattern observed, earlier works in bulk media made use of the effective three-wave mixing picture [15,25-26] to determine the region of the angularly-resolved spectra where the energy of the scattered waves gathers according to phase-matching constraints (similar to Eq. (4)). Briefly, the shape of this region depends on the linear chromatic dispersion of the medium. Its loci are determined by the velocity of the peak (here $\delta K_z^{(1)}$) in the nonlinearly-induced material response, whereas a finite bandwidth is induced by the interaction length over which nonlinear interactions are significant (here related to $\delta K_z^{(0)}$, and treated as a free parameter to describe the shape of the low-value contours of X-shaped pattern).

In Fig. 2a, we retrieve the same signatures of nonlinearly-generated conical waves, namely the phase-matching is satisfied over a certain spectral range for each spatial mode around a central frequency. The detailed nonlinear propagation for power spectrum, time and space profiles are shown in Figs. 2(b)-2(d), respectively. We observe that the spectral dynamics reaches a stationary state after 1.2 cm (see Fig. 2(b)), just past the pulse splitting phenomenon observed in the time domain. We then globally retrieve a typical scenario already analyzed in normal dispersion of bulk media, namely, the self-focusing dynamics and spectral broadening are associated to pulse splitting phenomena [5,27]. The pulse splitting occurring at the nonlinear spatial focus produces two sub-pulses moving in opposite directions in our retarded time frame (see Fig. 2(c)). As depicted in Fig. 2(e-f), we notice a strong and rapid spectral broadening at 1 cm induced by the self-steepening effect. In particular, the velocity difference between the trailing pulse peak and its tails leads to the formation of an optical shock (at the trailing edge) due to the nonlinear dependence of the refractive index. The shock front on the trailing edge is then followed by wave-breaking.

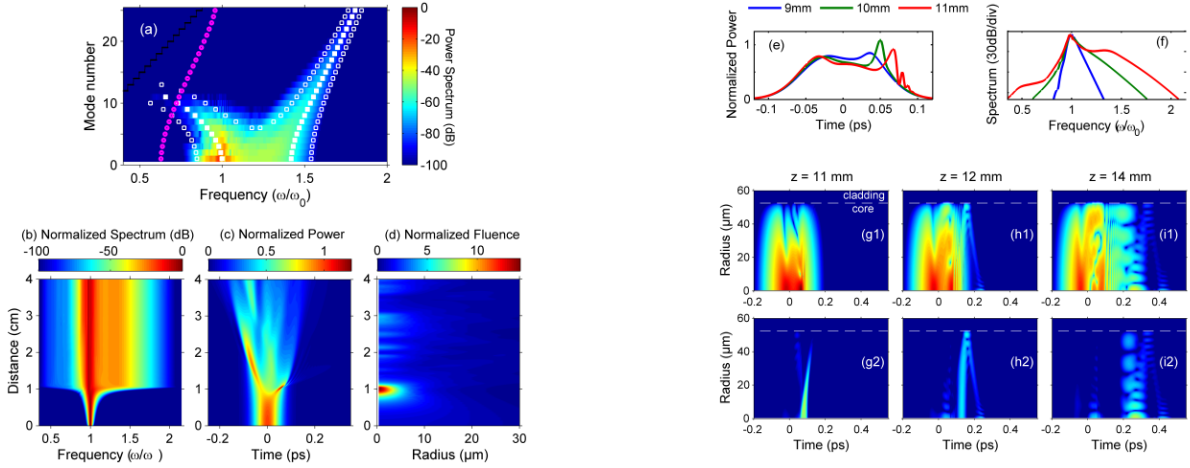

FIG. 2. Nonlinear propagation of 100-fs pulses (300-nJ energy at 800 nm) in a 4-cm-long segment of step-index MMF. (a) Output modal distribution of the power spectrum. White squares: theoretical phase-matching region of the discretized conical wave. Black curve: number of guided modes. Magenta circles: zero-dispersion frequency for each guided mode. (b-d) Evolution of the normalized full power spectrum, instantaneous power, and fluence with propagation distance, respectively. (e-f) Snapshots of temporal and spectral power profiles showing the optical shock formation and strong spectral broadening. (g1-i1) Evolution of the full spatiotemporal power distribution (over 6 decades) for distinct propagation distances. (g2-i2) Corresponding evolution for the filtered tails of X-pattern from $(m, \omega)$ plane.

The simultaneous strong broadening seeds linear waves, which are resonantly amplified in higher order modes according to the velocity of the shock front. This simply corresponds to phase-matched resonant radiations over the fiber modes (i.e., a discretized X-wave is formed), mainly satisfying the group velocity of the shock front as described by our relation (4). More specifically, $\delta K_z^{(1)} = 22$ ps m$^{-1}$, it can be found through the group velocity difference between the shock front and $1/K_{z(1,\omega_0)}^{(1)}$ in Fig. 2c, whereas $\delta K_z^{(0)} = \pm 10$ rad mm$^{-1}$ defines the finite bandwidth shown in Fig. 1a of our theoretical phase-matching (see open squares), and it is governed by the nonlinear contribution to the phase velocity. Note that a less intense optical shock on the leading edge of the pump pulse could be analyzed at 2 cm, seeding a less-visible discretized conical wave.

Next, we unveil the spontaneous formation of the conical wave in the direct $(r, t)$ space, in particular at the trailing edge (see Fig. 2(g1-i1)). A simple spectral filtering procedure of both tails of the X pattern from Fig. 2(a) allowed us to reveal its location and shape in the $(r, t)$ space (see Fig. 2(g2-i2)). We clearly retrieve the spatiotemporal signatures of the theoretical solution (9) depicted in Fig. 1(c-e), here the cone is emitted at the shock front position and then propagating into the fiber core in a dispersive manner. Note that various shapes of conical waves have been classified for bulk media as a function of the dispersive properties [28-29].

Overall, the dynamics above-observed in MMFs recalls the interplay between shock front (axial spectral broadening) and X-wave formation (conical spectral broadening) in a similar way to shocked-X wave dynamics described in femtosecond filamentation [30]. Moreover, it also makes the connection with well-known dynamics of dispersive shock waves studied in single-mode fibers which results in the emission of resonant radiations by the same phase-matching rule as our relation (4), but restricted to the fundamental mode [31]. The condition (4) can be seen as a generalized phase-matching rule of resonant radiations emitted by a broadband localized structure (soliton, shock front) through intra- and inter-modal properties of any waveguide and whatever the dispersion regime. We generalize the concept of conical waves towards structured media, where bulk media and single-mode waveguides become two limiting cases.

To this regard, we next investigate nonlinear propagation in the anomalous dispersion regime of the MMF, namely by pumping at $\lambda_0 = 1400$ nm. Figure 3(a) depicts the corresponding distribution of the full optical spectrum (power in log. scale) over the different fiber modes after 4 cm of propagation. A combination of both X-wave and O-wave is obtained in the $(m, \omega)$ plane, also known as a Fish-wave pattern. This spontaneous reshaping results from the large spectral broadening that covers both normal and anomalous dispersion regimes. Again, we are able to retrieve the corresponding discretized conical wave (by using the theoretical relation (4)), which is characterized by only one tail of the X-wave at high frequencies and the O-wave at low frequencies. If looking at the propagation details for spatial, temporal, and spatial profiles (not presented here), we would note a main pulse splitting phenomenon leading to the half X-wave pattern. We also checked that pumping farther from the zero dispersion (e.g., at $\lambda_0 = 1800$ nm) would result in the spontaneous O-wave formation (i.e. light bullet) [19,32].

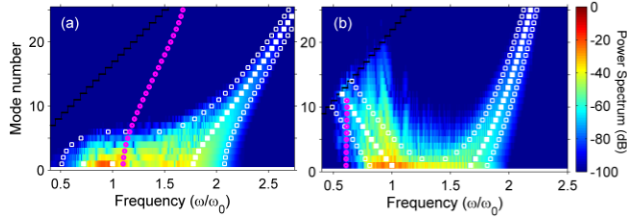

FIG. 3. (a) Same step-index MMF as in Fig. 1 with 450-nJ 100-fs pumping at 1400 nm. Output modal distribution of the power spectrum after 4 cm of propagation. (b) Output modal distribution of the power spectrum obtained from nonlinear propagation of 100-fs pulses (210-nJ energy at 800 nm) coupled into the $LP_{0,1}$ mode of a 2-cm-long GRIN fiber.

In addition, we present numerical results obtained in a 2-cm-long segment of gradient-index (GRIN) MMF pumped at $\lambda_0 = 800$ nm (see Fig. 3(b)). We consider a typical $GeO_2$-doped core with a diameter of 105-µm and a parabolic index profile, as well as a numerical aperture $NA = 0.2$. Figure 3(b) shows an evident X-shaped pattern of the field distribution over guided modes, but slightly different from the case of step-index MMFs. In particular, we observe distinct curvatures of conical tails spreading in higher-order modes. This specific behavior obtained in GRIN fibers is predicted by theoretical relation (4) of discretized conical waves (see white squares). The evolution of spectral, temporal and spatial profiles with propagation (not presented here) indicates the saturation of the spectral broadening after 5 mm, just past the first successive and periodic pulse splitting phenomena observed in the time domain associated to periodic self-imaging of GRIN fibers [33].

For the sake of clarity, we have here presented the fundamental mode excitation with femtosecond pulses. However, this is not a prerequisite for observing the spontaneous emergence of conical waves in MMFs. We point out that supercontinuum studies in single-mode fibers have demonstrated that the long-pulse pumping regime also lead to the emergence of ultrashort pulses and dispersive waves through an intermediate stage of modulation instability (MI) [8]. In the case of MMFs, various forms of MI can be observed [10] and considered as intermediate nonlinear processes possibly seeding a set of resonant radiations forming a discretized conical wave. Besides, a simple multimode excitation, by means of a large Gaussian beam, would imply that a part of the pump energy might remain at the carrier frequency for most of the higher order modes excited. This typical pumping configuration was studied experimentally in several other works [34-36], in particular for GRIN fibers. Such fibers exhibit well-confined fundamental guided modes in strong contrast to step-index fibers, the latter class of MMFs then appears as better candidates for future experimental demonstration of discretized conical wave formation.

*Conclusion.* In summary, we have unveiled the existence of discretized conical waves in structured media, and more particularly multimode fibers. Such space-time wave packets with arbitrary group-velocity result from a linear superposition of fiber modes with an engineered spatiotemporal spectrum. Moreover, they can spontaneously emerge during nonlinear propagation of ultrashort pulses similarly to bulk conical emission. The multimode fiber is an exciting medium, combining bulk and fiber properties to revisit the physics of pulse propagation with new degrees of freedom, as already revealed in planar glass membrane fibers [37].

*Acknowledgments.* We acknowledge financial support of the French "Investissements d'Avenir" program (PIA2/ISITE-BFC, contract ANR-15-IDEX-03; EIPHI Graduate School, contract ANR-17-EURE-0002). The authors thank K. Tarnowski and S. Majchrowska for stimulating discussions. Calculations were performed using HPC resources from DNUM-CCUB (Université de Bourgogne).

---

* pierre.bejot@u-bourgogne.fr